# Random Residue Sequences and the Number Theoretic Hilbert Transform

Vamsi Sashank Kotagiri
Oklahoma State University, Stillwater

**Abstract**
This paper presents random residue sequences derived from the number theoretic Hilbert (NHT) transform and their correlation properties. The autocorrelation of a NHT derived sequence is zero for all non-zero shifts which illustrates that these are self-orthogonal sequences. The cross correlation function between two sequences may be computed with respect to the moduli of the either sequence. There appears to be some kind of an inverse qualitative relationship between these two different computations for many sets of residue sequences.

**Introduction**
The search for random sequences with ideal randomness properties is an important area of computer science. Shift register sequences provide near-ideal autocorrelation function but the period is constrained to be $2^n-1$, for different values of n [1]. Another interesting random sequence family is that of "decimal" sequences [2]-[6], which is of particular interest since any sequence can be described also as a "decimal sequence". Here we wish to go beyond the results of these sequences by obtaining residue sequences modulo prime that have ideal autocorrelation function, that is it is zero for all non-zero values of the argument. In other words, rather than binary sequences, we wish to deal with sequences where individual items are integers modulo a prime. We do so by using the structure of the NHT matrix.

The NHT, introduced recently [7]-[9] as a generalization of the discrete Hilbert transform (DHT) [10], is a circulant matrix with alternating entries of each row being zero and non-zero numbers and transpose modulo a suitable number is its inverse. The DHT has found many applications in signal processing and also in scrambling and in a variety of other applications in speech and image analysis [11]-[16] and it is closely related to scrambling transformations [17]-[19].

The idea of the use of the NHT matrix to generate random residue sequences comes from the fact that the product of the NHT matrix with its transpose computes all correlations on the block. Therefore the circulant part of the NHT matrix should be able to generate ideal random sequences. Since the 32-point NHT has only 16 non-zero values, we will generate 16-point random sequences and plot its autocorrelation function. Furthermore, we wish to explore the cross correlation properties of such sequences. For two sequences, the cross correlation can be taken with respect to the modulus of either sequence.

**Background**

In earlier papers [8],[9] NHT matrices were presented for up to 16-points. Let us consider the data block to be F and the NHT transform to be N. The matrix N is the general form of NHT and m is appropriate value of modulus; its inverse is $N^T$ mod q.

For a block of data F, the NHT transform $G = NF$ mod $q$. The inverse of the transform is F = $N^T$G mod q. In other words,

$$NN^T = I \tag{1}$$

where I is the identity matrix. We represent the first row of matrix by using integers *a,b,c,d,e,f,g,h,i,j,k,l, m,n,o,p* and alternate with 0s. When we multiply this circulant matrix with its transpose we observe that the squares of the integer values of the first row assumed above will be equal to 1 modulo the chosen q. In other words,

$$a^2 + b^2 + c^2 + d^2 + e^2 + f^2 + g^2 + h^2 + i^2 + j^2 + k^2 + l^2 + m^2 + n^2 + o^2 + p^2 \tag{2}$$

The other non-diagonal element terms in the product $NN^T$ are:

$$2(ai + bj + ck + dl + em + fn + go + hp) \tag{3}$$
$$a(h+j)+b(i+k)+c(j+l)+d(k+m)+e(l+n)+f(m+o)+g(n+p)+pi+oh \tag{4}$$
$$a(g+k)+b(h+l)+c(i+m)+d(j+n)+e(k+o)+f(l+p)+gm+hn+io+jp \tag{5}$$
$$a(f+l)+b(g+m)+c(h+n)+d(i+o)+e(j+p)+k(f+p)+gl+hm+in+jo \tag{6}$$
$$a(e+m)+b(f+n)+c(g+o)+d(h+p)+i(e+m)+j(f+n)+k(g+o)+l(h+p) \tag{7}$$
$$a(d+n)+b(e+o)+c(f+p)+g(d+j)+h(e+k)+i(f+l)+m(j+p)+kn+lo \tag{8}$$
$$a(c+o)+b(d+p)+e(c+g)+f(d+h)+i(g+k)+j(h+l)+m(k+o)+n(l+p) \tag{9}$$
$$ab + bc + cd + de + ef + fg + gh + hi + ij + jk + kl + lm + mn + no + op + pa. \tag{10}$$

The correlations amongst the first 8 elements are mirrored to the remaining 8 elements therefore 8 conditions for non diagonal elements are shown. The modulus should be chosen in such a way that the term (2) is 1 and the terms (3)-(10) are zeros. This provides constraints on what can be chosen to be the modulus. The autocorrelation function captures the correlation of data with itself. For a data sequence *a(n)* of N points the autocorrelation function *C(k)* is represented by

$$C_a(k) = \frac{1}{N}\sum_{j=1}^{N} a(j)a(j+k) \tag{11}$$

For a noise sequence, the autocorrelation function $C_a(k) = E(a(i)a(i+k))$ is two-valued, with value of 1 for k=0 and a value approaching zero for k≠0 for a zero-mean random variable. Assuming periodicity, such a sequence will have C(k) as 1 for k=0 and approximately $\mu^2$ for non-zero *k*. $\mu$ is the mean of the variable.

We now present an algorithm for generating the NHT generator sequence. The basic idea that has worked very well is to pick a number that is prime and then pick numbers that are powers of 2.

**Algorithm for Generating the Sequence**
1. Enter the number of rows and columns of the circulant matrix generally the number of rows and columns will be equal to desired NHT .i.e. (16 by 16 for 16-point NHT, 12 by 12 for 12-point NHT).
2. Then choose the elements of circulant NHT matrix in such a way that one element need to be a prime number and the remaining elements need to be 2 and multiples of 2.i.e (7 2 4 8 16 32 for 12-point NHT).Circulant NHT matrix form can be seen in [13].
3. Then find the transpose of the circulant NHT matrix.
4. Multiply the circulant NHT matrix with its transpose matrix and we will get a product NHT matrix.
5. Find the gcd of all the non-diagonal elements in the obtained product NHT matrix.
6. The gcd of non-diagonal elements will be the modulus of the circulant NHT matrix, in most of the cases it will be a prime modulus.
7. If the desired format is I mod n, the we need to normalize the elements of the NHT matrix with the remainder obtained by taking modulus of the diagonal elements.
8. We will get the elements **$a_1$, $a_2$, $a_3$ …..$a_n$** and modulus which will be a prime number.

*Pseudocode of the algorithm*
1. Input rows[m] and columns[n] of matrix .i.e matrix[m][n]
2. for i<-0 to m do                                  // Input circulant matrix
3. for j<-0 to n do
4.  Matrix[i][j] <- values.
5. for i<-0 to m do
6. for j<-0 to n do                                  //Finding transpose of matrix
7. transpose[j][i]=matrix[i][j]
8. for i<-0 to n do
9. for j<-0 to m do
10. transpose[i][j]
11. for i<-0 to m do
12. for j<-0 to n do
13. for k<-0 to m do                                 //Multiply circulant with its transpose
14. sum <-0
15. sum = sum + matrix[i][k]*transpose[k][j];
16. mul[i][j]=sum;
17. for i<-0 to 1 do                                 //Initialize non diagonal elements of matrix
18. for j<-1 to n do
19. mul[i][j];
20. gcd=greatestcommondivisor(gcd,mul[i][j]);
21. greatestcommondivisor( a , b )                   //gcd of non diagonal elements.
22. while ( a % b != 0) do
23. x=b;
24. y=a%b;
25. a=x;
26. b=y;
27. return b;

**Experimental Results for Autocorrelation Function**

We have done random experiments by using the above algorithm for different values of input sequences those are represented in the below table. As seen from Figures 1-4, their amplitudes have a variety of relationships and their autocorrelation functions have been plotted.

Table 1. Four example 16-bit long NHT sequences

| Example | a | b | c | d | e | f | g | h | i |
|---|---|---|---|---|---|---|---|---|---|
| 1 | 911 | 1821 | 3642 | 1 | 2 | 4 | 8 | 16 | 32 |
| 2 | 12747 | 3642 | 7284 | 14568 | 7285 | 14570 | 7289 | 14578 | 7305 |
| 3 | 3 | 2 | 4 | 8 | 16 | 32 | 64 | 128 | 256 |
| 4 | 2 | 2 | 4 | 8 | 16 | 32 | 64 | 128 | 256 |
| 5 | 11 | 2 | 4 | 8 | 16 | 32 | 17 | 34 | 21 |
| 6 | 13 | 2 | 4 | 8 | 16 | 32 | 64 | 128 | 256 |

| j | k | l | m | n | o | p | mod q |
|---|---|---|---|---|---|---|---|
| 64 | 128 | 256 | 512 | 1024 | 2048 | 4096 | 7283 |
| 14610 | 7369 | 14738 | 7625 | 15250 | 8649 | 17298 | 21851 |
| 512 | 1024 | 2048 | 975 | 1950 | 779 | 1558 | 3121 |
| 181 | 31 | 62 | 124 | 248 | 165 | 330 | 331 |
| 42 | 37 | 27 | 7 | 14 | 28 | 9 | 47 |
| 512 | 1024 | 61 | 122 | 244 | 488 | 976 | 1987 |

Random sequences from the first four examples of Table 1, when plotted on a graph, will be as follows.

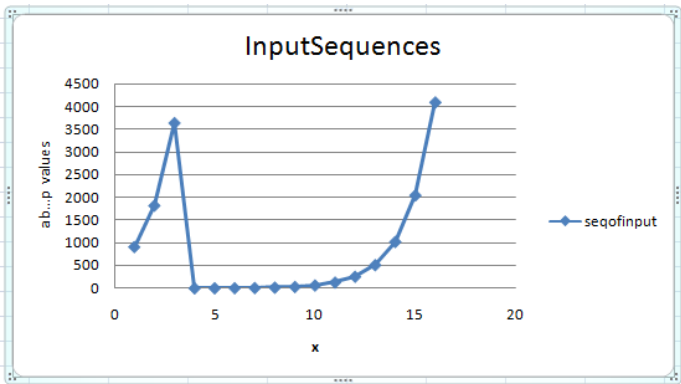

Figure 1: Input graph for the above values

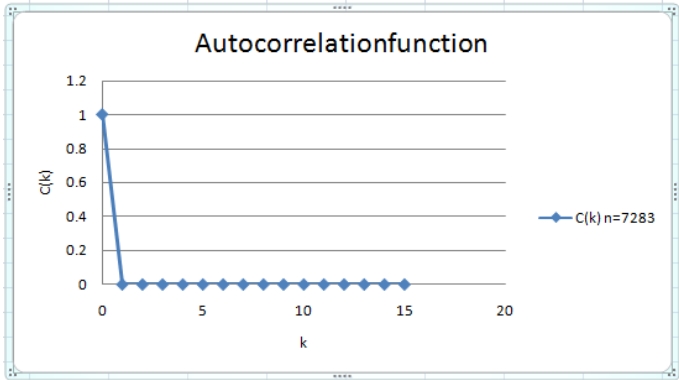
Figure 2: Autocorrelation function for the above input

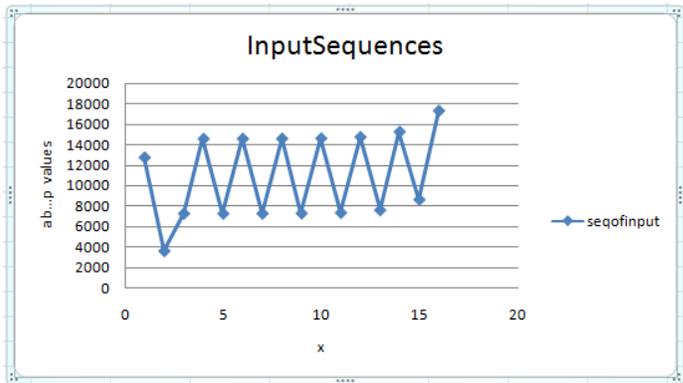
Figure 3: Input graph for the example 2 values mod 21851

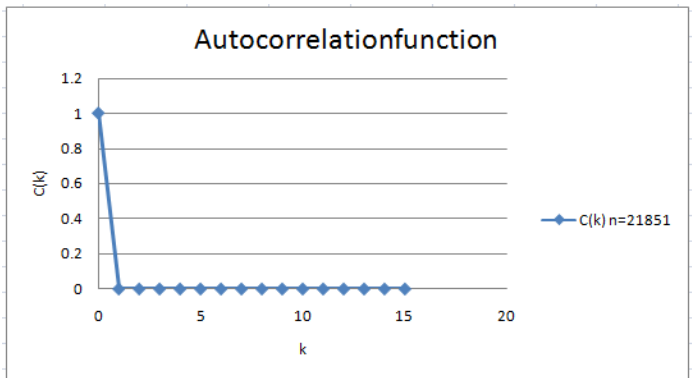
Figure 4: Autocorrelation function

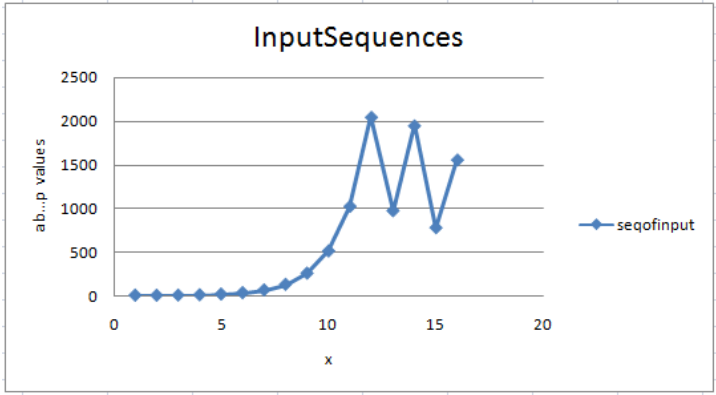

Figure 5: Input graph for example 3 values mod 3121

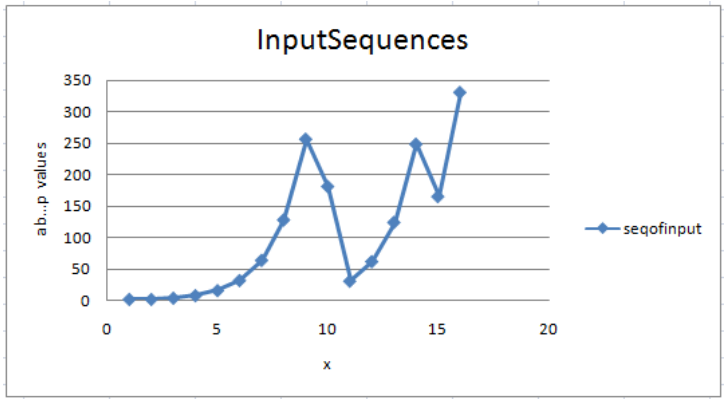

Figure 6: Input graph for example 4 values mod 331

As is to be expected the autocorrelation function function $C_a(k)$ is zero for all values where $k \neq 0$. This is also true for examples 3 and 4 for which the autocorrelation graph is not included for reasons of economy. The autocorrelation property makes the sequences very attractive for a variety of signal processing applications.

**Cross-correlation function**
The cross correlation function captures the correlation of data with other sequences. For a data sequence $a(n)$ of $N$ points the cross correlation function $C(k)$ is represented by

$$C_c(k) = \frac{1}{N} \sum_{j=1}^{N} a(j)b(j+k) \qquad (11)$$

where the value b (j +k) corresponds to the next sequence result.

For a noise sequence, the cross correlation function $C_c(k) = E(a(i)b(i+k))$. If the two sequences are independent the cross correlation will be the product of their individual means.

## Experimental Results for Cross-Correlation Function

We continue with the first four examples of Table 1 and we compute their mutual cross-correlation functions.

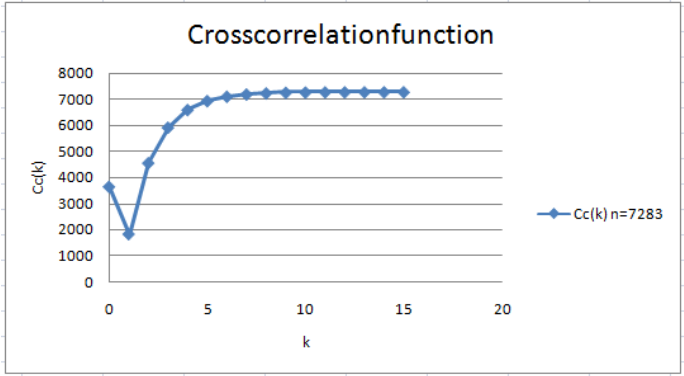

Figure 7: Cross correlation function between example 1 and 2 when mod n=7283

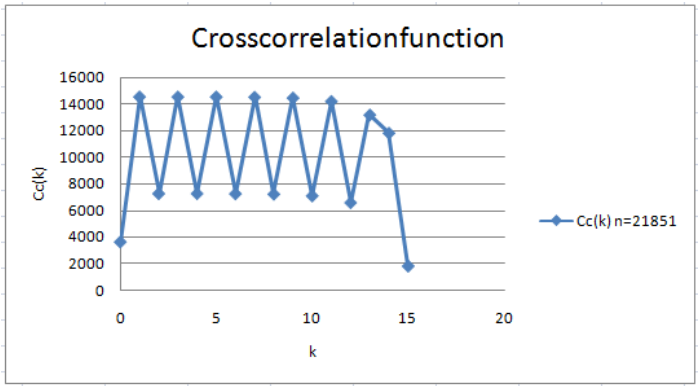

Figure 8: Cross correlation function between example 1 and 2 when mod n=21851

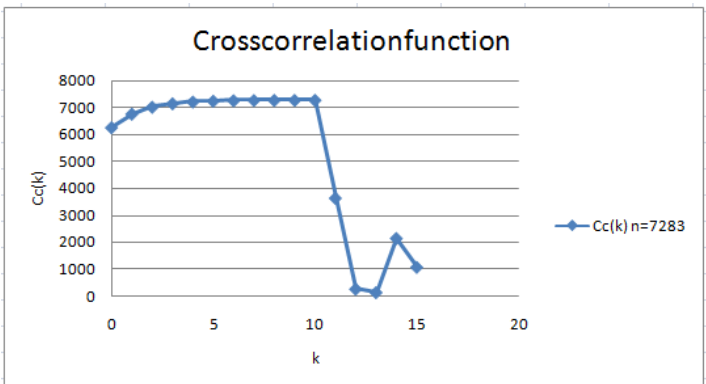

Figure 9: Cross correlation function between example 1 and 3 when mod n=7283

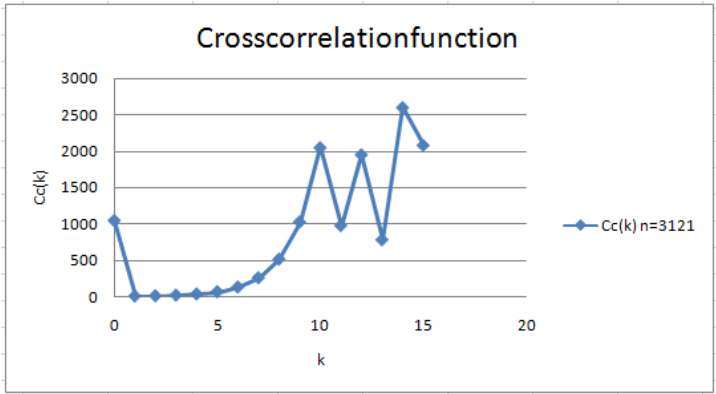

Figure 10: Cross correlation function between example 1 and 3 when mod n=3121

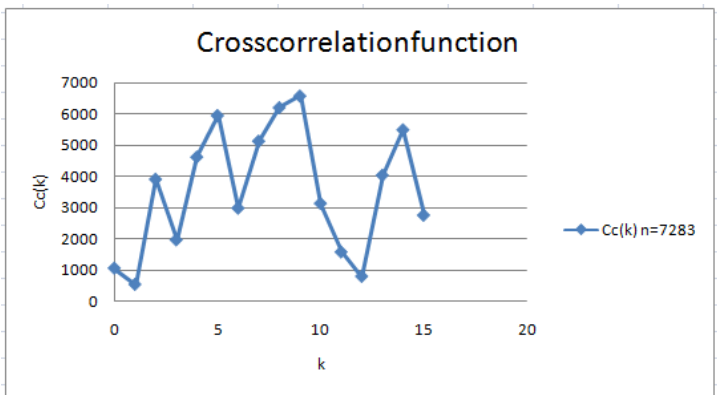

Figure 11: Cross correlation function between example 1 and 4 when mod n=7283

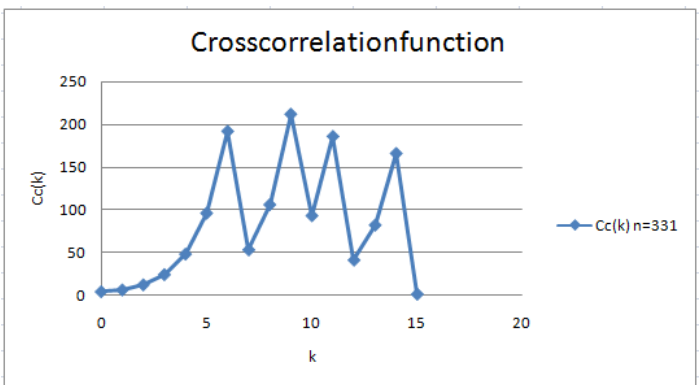

Figure 12: Cross correlation function between example 1 and 4 when mod n=331

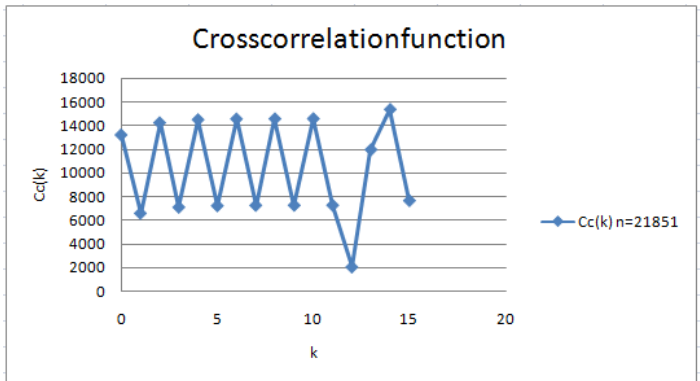

Figure 13: Cross correlation function between example 2 and 3 when mod n=21851

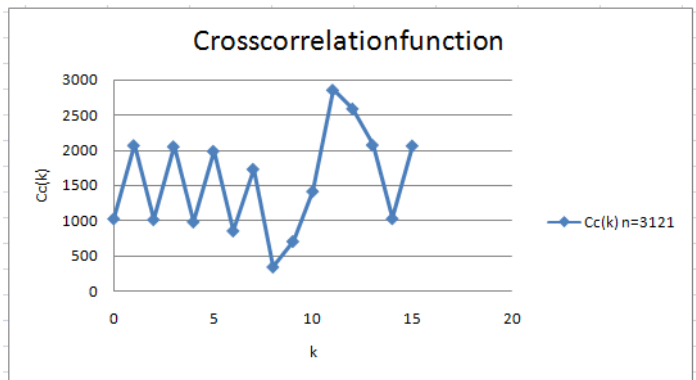

Figure 14: Cross correlation function between example 2 and 3 when mod n=3121

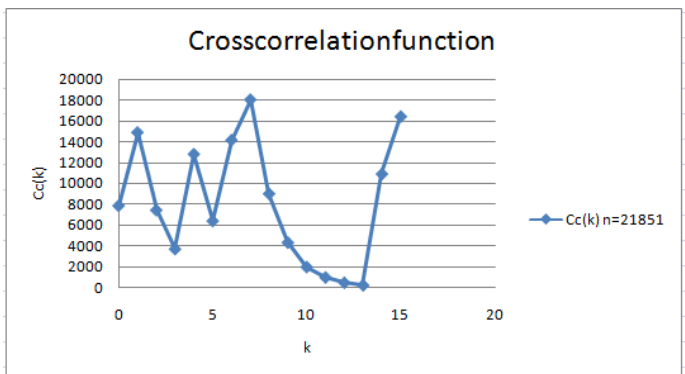

Figure 15: Cross correlation function between example 2 and 4 when mod n=21851

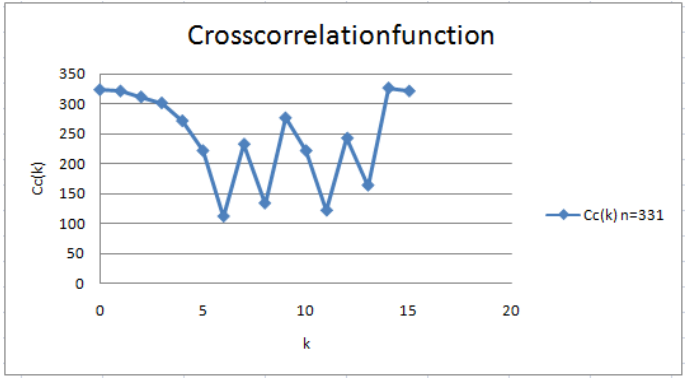

Figure 16: Cross correlation function between example 2 and 4 when mod n=331

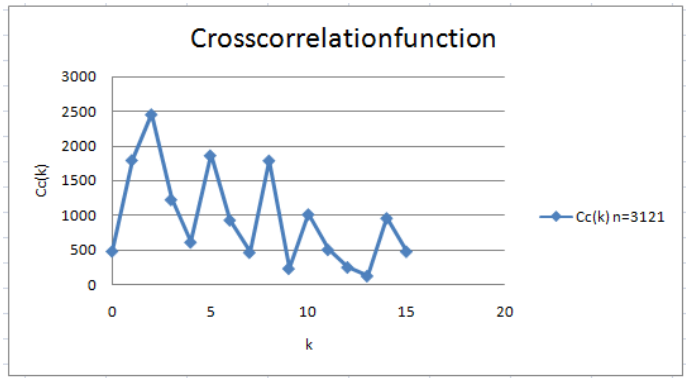

Figure 17: Cross correlation function between example 3 and 4 when mod n=3121

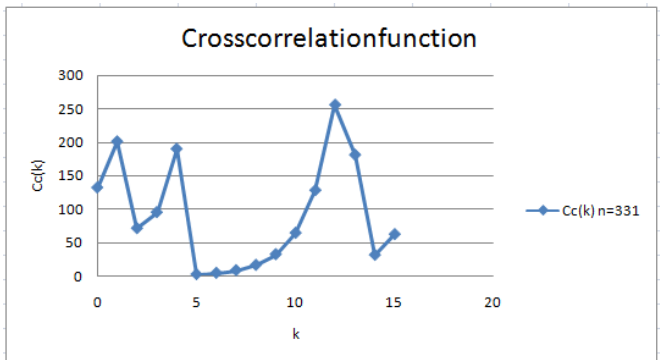

Figure 18: Cross correlation function between example 3 and 4 when mod n=331

In order to compare the different cross correlation values amongst sequences with respect to different moduli, we propose the use of the following measure:

$$E(C_c) = \frac{\sum_k C_c(k)}{period \times n} \quad (12)$$

The cross correlation between examples i and j will be represented by:

$$C_c^{ij}(k) \mod n_i \tag{13}$$

where it is understood that $n_i$ is the modulus for example $i$. The results are presented in Table 2.

Table 2. Expectation of cross correlation

| Case | Sequences i and j | Expectation of Cross correlation |
|---|---|---|
| 1 | 1 and 2 | 0.87 |
| 2 | 2 and 1 | 0.45 |
| 3 | 1 and 3 | 0.73 |
| 4 | 3 and 1 | 0.27 |
| 5 | 1 and 4 | 0.49 |
| 6 | 4 and 1 | 0.25 |
| 7 | 2 and 3 | 0.47 |
| 8 | 3 and 2 | 0.50 |
| 9 | 2 and 4 | 0.37 |
| 10 | 4 and 2 | 0.74 |
| 11 | 3 and 4 | 0.31 |
| 12 | 4 and 3 | 0.28 |

It is interesting that in half the cases the expectation of the cross correlation is complementary across the two moduli. For example, cases 3 and 4 have average cross correlation of 0.73 and 0.27, respectively. Likewise, cases 7 and 8, and cases 9 and 10 are complementary.

**Conclusions**

This paper presented several sequences of length 16 with entries defined modulo a prime that have perfect autocorrelation properties and variable cross correlation properties. The fact of zero autocorrelation function for all non-zero lags makes them suitable for use in applications where orthogonal sequences are needed.

Several example sequences have been investigated for their cross correlation properties and it is found that these appear to have an inverse relationship for the two moduli that may have cryptographic applications. An interesting problem to investigate will be finding classes of pairs of sequences that have average complementary cross correlation.